\documentclass[12pt]{article}
\newif\ifams\amsfalse                                                    
\amstrue                                                                 
\ifams                                                                   
 \usepackage{amsfonts}                                                   
\fi                                                                      
\newif\iffigs\figsfalse                                                  
\figstrue                                                               
\newif\ifdraft\draftfalse
\newif\ifinter\interfalse
\intertrue
\ifdraft\else\interfalse\fi
\ifdraft
 \else
\fi
\ifinter
  \def\secl#1{\nopagebreak\marginpar{\vspace{-6mm}\scriptsize #1}\label{#1}}
  \def\beql#1{\marginpar{\vspace{4mm}\scriptsize #1}
              \nopagebreak\begin{equation}\label{#1}}
  \def\ftl#1#2{\footnote{\label{#1}[#1] #2}}
  \def\capl#1#2{\caption{[#1] #2}\label{#1}}
  \def\bibl#1{\marginpar{\vspace{4mm}\scriptsize #1}\nopagebreak\bibitem{#1}}
 \else
  \def\secl#1{\label{#1}}
  \def\beql#1{\begin{equation}\label{#1}}
  \def\ftl#1#2{\footnote{\label{#1}#2}}
  \def\capl#1#2{\caption{#2}\label{#1}}
  \def\bibl#1{\bibitem{#1}}
\fi

\def\tempnote#1%
  {\hspace{1mm}\newline\noindent\begin{tabular}[t]{|p{15.5cm}|}
     \hline \rule{0mm}{2.5ex}#1 \\ \hline
    \end{tabular}\newline\noindent}

\def\draftnote#1%
  {\ifdraft
    \hspace{1mm}\newline\noindent\begin{tabular}[t]{|p{15.5cm}|}
     \hline \rule{0mm}{2.5ex} \underline{DRAFT NOTE}: #1 \\ \hline
    \end{tabular}\newline\noindent
   \fi}

\def\internote#1%
  {\ifinter
    \hspace{1mm}\newline\noindent\begin{tabular}[t]{|p{15.5cm}|}
     \hline \rule{0mm}{2.5ex} \underline{Internal Note}: #1 \\ \hline
    \end{tabular}\newline\noindent
   \fi}

\def\multdn{$\downarrow\downarrow\downarrow\downarrow\downarrow$}
\def\beginsup%
 {\noindent\begin{tabular}[t]{|c|}
   \hline  Supplements\\
   \multdn\multdn\multdn\multdn\multdn\multdn
   \multdn\multdn\multdn\multdn\multdn\multdn \\ \hline
  \end{tabular}}

\def\multup{$\uparrow\uparrow\uparrow\uparrow\uparrow$}
\def\endsup%
 {\noindent\begin{tabular}[t]{|c|}
   \hline \multup\multup\multup\multup\multup\multup
          \multup\multup\multup\multup\multup\multup \\
   Supplements\\ \hline
  \end{tabular}}

\iffigs
 \input epsf
 \def\pct#1{\centerline{ \epsfbox{#1.eps}}}
\else
 \def\pct#1{(see figure in file #1.eps)}
\fi
\newif\ifappend\appendfalse
\appendtrue
\ifappend
 \newcommand{\newsection}[1]{
  \vspace{7mm} \pagebreak[3]
  \refstepcounter{section}
  \setcounter{equation}{0}
  \message{(\thesection. #1)}
  \addcontentsline{toc}{section}{
   \protect\numberline{\thesection}{\hs\hs\boldmath #1}}
  \begin{flushleft}
   {\large\bf\boldmath \thesection. #1}
  \end{flushleft}
  \nopagebreak}

\else
 \newcommand{\newsection}[1]{\section{#1}}
\fi

\def\al{\alpha}
\def\bt{\beta}
                \def\Gm{\Gamma}
\def\dl{\delta}                \def\Dl{\Delta}
\def\ep{\epsilon}

\def\lm{\lambda}               

\def\th{\theta}               
\def\vph{\varphi}

\def\om{\omega}               \def\Om{\Omega}
\def\sg{\sigma}               \def\Sg{\Sigma}
\def\zt{\zeta}

\def\Ac{\mbox{\protect$\cal A$}}

\def\Fc{\mbox{\protect$\cal F$}}

\def\Mc{\mbox{\protect$\cal M$}}

\def\Oc{\mbox{\protect$\cal O$}}

\ifams
 \def\bbl#1{{\mathbb #1}}
\else
 \def\bbl#1{{\bf #1}}
\fi

\def\RR{\bbl{R}}
\def\ZZ{\bbl{Z}}

\def\diag{{\rm diag}}

\def\beq{\begin{equation}}
\def\eeq{\end{equation}}
\def\hs{\hspace{2mm}}
\def\hsc{\hspace{2mm},\hspace{5mm}}
\def\nl{\protect\newline}
\def\nlb{\protect\newline $\bullet$ }
\def\nls{\protect\newline $\star$ }
\def\ie{{\em i.e.}}

\def\pt{\partial}
\def\goto{\rightarrow}
\def\Goto{\hspace{5mm}\Rightarrow\hspace{5mm}}

\def\wbar{\overline}
\def\what{\widehat}
\def\dg{^{\dagger}}
\def\inv{^{-1}}
\def\siml{\raisebox{-1ex}{$\stackrel{\textstyle <}{\sim}$}}

\def\rec#1{{\raise 0.4ex \hbox{$\scriptstyle {\frac{1}{#1}}$}}}

\def\half{{\raise 0.4ex \hbox{$\scriptstyle {1 \over 2}$}}}

\def\hepth#1{{\tt hep-th/#1}}

\def\JGP#1#2#3{{\it J. Geom. Phys.} {\bf #1} (#2) #3} 
\def\JHEP#1#2#3{{\it JHEP} {\bf #1} (#2) #3}
 
\def\NPB#1#2#3{{\it Nucl. Phys.} {\bf B#1} (#2) #3}

\def\PRV#1#2#3{{\it Phys. Rev.} {\bf #1} (#2) #3}

\newif\iftpage\tpagefalse                                               
\tpagetrue                                                             
\begin{document}

\iftpage
 \begin{titlepage}
 \ifdraft
   \fbox{
   \ifinter INTERNAL \fi
   DRAFT}\vspace{-1cm}
 \fi
\else
 \ifdraft
  \pagestyle{myheadings}
  \markright{\fbox{
  \ifinter INTERNAL \fi
  DRAFT}}
 \fi
\fi

\begin{flushright}
EFI-2000-23\\ {\tt hep-th/0007100}\\
\end{flushright}

\ifinter \vspace{-10mm} \else \vspace{5mm} \fi

\begin{center}
\LARGE {\bf\boldmath On the Quantization Constraints \\
for a D3 Brane \\ in the Geometry of NS5 Branes} \\

\ifinter \vspace{5mm} \else \vspace{10mm} \fi

\large Oskar Pelc \normalsize 
\vspace{5mm}

{\em Enrico Fermi Institute\\ University of Chicago \\
5640 S. Ellis Ave. Chicago, IL 60037, USA} \\
E-mail: {\tt oskar@theory.uchicago.edu}

\vspace{5mm}
\ifinter \else \vspace{5mm} \fi
\end{center}

\ifinter \else \vspace{10mm} \fi

\begin{center}\bf Abstract\end{center}
\begin{quote}
A D3 brane in the background of NS5 branes is studied semi-classically.
The conditions for preserved supersymmetry are derived, leading to a
differential equation for the shape of the D3 brane. The solutions of
this equation are analyzed. For a D3 brane intersecting the NS5 branes,
the angle of approach is known to be restricted to discrete values.
Four different ways to obtain this quantization are described.
In particular, it is shown that,
assuming the D3 brane avoids intersecting a {\em single} NS5 brane,
the above discrete values correspond to the different possible positions of
the D3 branes among the NS5 branes.
\end{quote}


%

\iftpage
 \end{titlepage}
 \ifdraft
  \pagestyle{myheadings}
  \markright{\fbox{
  \ifinter INTERNAL \fi
  DRAFT}}
 \fi
\fi

\tableofcontents


\newsection{Introduction and Summary}
The dynamics of branes in string theory is an important field of active
research, which is hoped to provide clues about the fundamental description
of the theory.
There are several approaches to this subject, leading to complementary
information.
Treating the branes as classical manifolds, many results can be obtained
using duality and supersymmetry arguments.
Alternatively, the branes can be represented by the supergravity background
that they induce, providing an additional point of view.
A particularly fruitful approach is to represent the NS5 brane by its
induced geometry and to consider D branes in this background.
First, the NS5 geometry is relatively simple so, considering the D branes as
classical manifolds, the resulting equations for the shape of the brane are
tractable. Second, the Ramond-Ramond fields vanish in this geometry, so
the worldsheet formulation of string theory in this background is well
understood and can be used.
In this formulation, the D branes are represented by open
strings with specified boundary conditions.
Finally, the near-horizon region of the NS5 brane geometry corresponds to
a {\em solvable} worldsheet description, enabling the derivation of
exact results.

In this work we consider the following static brane configuration in type
IIB string theory (see figure \ref{f-cf}):
\begin{figure}
\pct{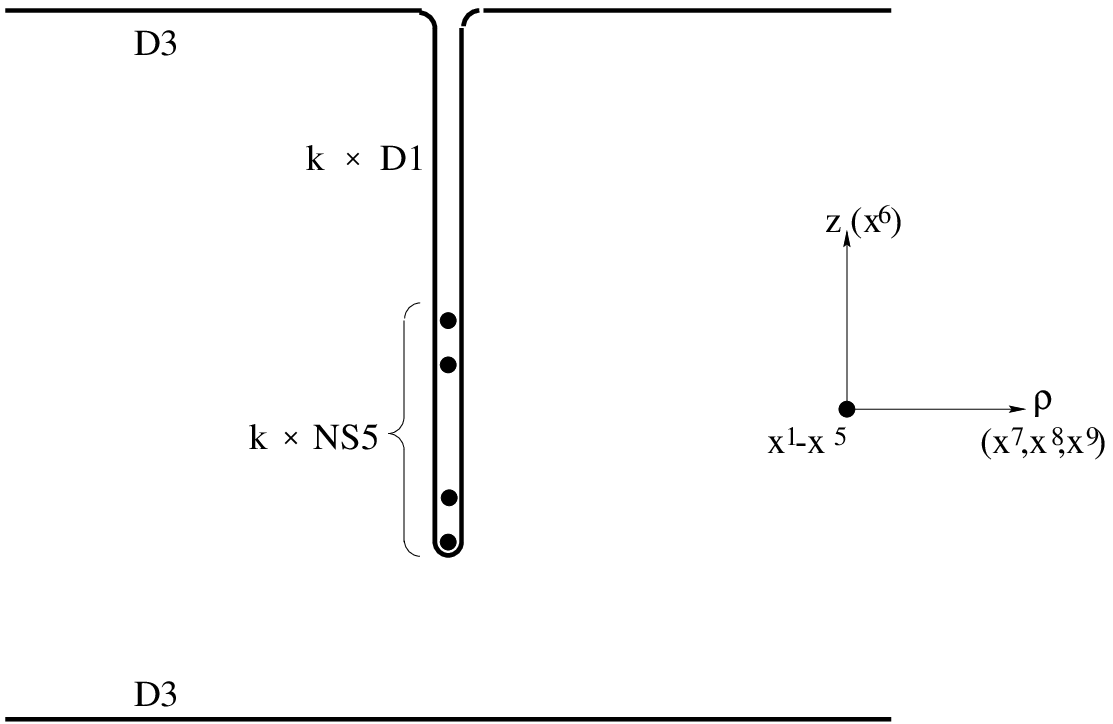}
\capl{f-cf}{The brane configuration}
\end{figure}
$k$ parallel NS5 branes, distributed along a line (parametrized by $z$)
and a D3 brane%
\footnote{Using T-duality in directions parallel to the NS5 branes, the D3
brane can be replaced by a D$(3+p)$ brane with $p$ direction parallel to the
NS5 branes, without influencing the results obtained here.}
orthogonal to the NS5 branes and to $\hat{z}$.
This configuration has an $SO(3)$ symmetry and preserves $\frac{1}{4}$ of
the 32 supercharges in the theory.
One can add a D1 brane along the $z$ axis without breaking the above
symmetries.
Moreover, by moving the D3 brane in the $z$ direction,
such a D1 brane is created as the D3 brane crosses an NS5 brane.
This is known as the Hanany-Witten effect \cite{HW9611}.
In the present treatment, the NS5 branes are represented by the geometry that
they induce \cite{CHS9112}, which is of the form
\[ \RR^{5,1}\times\Mc \]
and the D3 brane is considered as a classical 3-manifold $D$ in the
4-dimensional geometry $\Mc$ transverse to the NS5 branes.
Because of the $SO(3)$ symmetry, the shape of this manifold is described by a
single function $z(\rho)$.
The supersymmetry of this system is analyzed and it is shown
(following the approach in \cite{CM9708}\cite{Imamura9807})
that the preservation of supersymmetry translates to a restriction on the
shape of the brane, expressed by a differential equation for $z(\rho)$.
Analyzing the solutions of this equation, the Hanany-Witten effect
is reproduced: the D3 brane approaches, for $\rho\goto\infty$, a flat
hyper-surface at $z=z_\infty$, where $z_\infty$ is a continuous parameter;
starting with a flat D3 brane at $z=-\infty$ (see figure \ref{f-cf}), and
moving it ``up'' (increasing $z_\infty$), the D3 brane is repelled from
the NS5 branes and, as a result, a tube is formed;
when the D3 brane is high enough, the configuration looks for a
distant observer as a 1-brane along the $z$ axis extended between the
NS5 branes and a flat D3 brane; by calculating the charge and tension%
\ftl{f-T}{The tension is calculated using the Dirac-Born-Infeld action for
a D brane.}
of this 1-brane, it is identified as $k$ coinciding D1 branes.

The family of configurations described above is part of a larger family,
with an additional parameter $\psi_0$, which characterizes the 2-form
field $\Fc$ on the brane (the above sub-family corresponds to $\psi_0=\pi$).
Classically, this parameter can take any value, however,
when quantum considerations are taken into account,
one finds that only discrete values are allowed:
\beql{psi-disc} \psi_0=N\frac{\pi}{k} \hs. \eeq
Of particular interest are configurations in which all the $k$ NS5 branes
coincide and the D3 brane intersects (touches) them.
This corresponds to the range $0<\psi_0<\pi$ and in this case $\psi_0$ is
the angle at which the D3 brane approaches the NS5 branes. 
The restriction (\ref{psi-disc}) implies in this case that only $k-1$ values
are allowed for this angle, corresponding to $N=1,\ldots,k-1$.

Remarkably, there are four independent ways to derive the integrality
condition (\ref{psi-disc}):
\begin{enumerate}
\item Demanding that the coupling of the fundamental string to the 2-form
field $\Fc$ on the D brane is well defined;
\item Demanding that the number of D1 branes ending on the D3 brane
is integral;
\item Assuming that a D3 brane does not intersect a single NS5 brane;
\item Using the worldsheet description in the near-horizon region, where
exact quantum analysis can be performed.
\end{enumerate}
We describe the first approach in general (following \cite{KS9609})
and show that, in special situations, it translates to an integrality
condition
\beql{F-disc} \frac{1}{2\pi}\int_{C_2}F\in\ZZ \hsc \forall C_2\in H_2(D,\ZZ)
\eeq
on the gauge field strength $F$ on the D brane. This condition was proposed
recently in \cite{BDS0003} (and discussed further in
\cite{Pawel0003}-\cite{Stanciu0006})
and here this proposal is confirmed and clarified.
Applied to the present case, it leads to the condition (\ref{psi-disc}).
In the second approach, a D3 brane with $\psi_0>0$ is considered.
For $z_\infty$ large enough, such a configuration includes a tube connected
to the assympoticly flat part of the D3 brane (as illustrated, for
$\psi_0=\pi$, in figure \ref{f-cf}).
For $z_\infty\goto\infty$, the tube shrinks to zero width and it is shown to
have a charge and tension$^{\ref{f-T}}$ of $N=\frac{k\psi_0}{\pi}$ D1 branes.
Integrality of $N$ leads, again, to the condition (\ref{psi-disc}).

In both these approaches, the integrality condition (\ref{psi-disc}) is
derived from a consistency requirement and leaves the real origin of the
restriction obscure. The third approach provides a more natural starting
point. The NS5 branes are distributed along the $z$ axis, and it is shown
that an integer $N$ in eq. (\ref{psi-disc}) corresponds to a D3 brane passing
between the $N$th and the ($N+1$)th NS5 brane, while for $\hat{N}-1<N<\hat{N}$
(with $\hat{N}=1,\ldots,k$) the D3 brane intersects the $\hat{N}$th NS5 brane.
Thus, the restriction (\ref{psi-disc}) simply means that the D3 and NS5
branes avoid intersecting each-other and they do intersect only if this is
unavoidable, as is the case when a D3 brane is trapped between coinciding
NS5 branes.

Finally, in the forth approach, one restricts attention to 
the near-horizon region of coinciding NS5 branes and considers
the formulation of the string dynamics in this background in terms of
the corresponding 2D conformal field theory.
The D branes studied so far in this approach are described in spherical
coordinates by $\psi=\psi_0$ (\ie, $\psi$ independent of the radial
coordinate $r$). Note that this is only a subset of the configurations found
in the semi-classical approach described above:
$\psi_0$ is restricted to the range $(0,\pi)$ and, for each such $\psi_0$,
there is a single value of $z_\infty$.
Imposing the symmetry requirements, one obtains (following the approach
in \cite{Cardy89}) $k-1$ types of branes.
A calculation of
the expectation values of bulk fields \cite{FFFS9909} provides evidence
that these branes correspond to the values in eq. (\ref{psi-disc})
with $N=1,\ldots,k-1$.
It is worth noting that there are no additional branes corresponding to
$\psi_0=0,\pi$, \ie, to D1 branes. Instead, D1 branes are seen in the
near-horizon region as cylindrical D3 brane. For a single D1 brane, the
radius of this cylinder is comparable or smaller to the string scale, so it
indeed should be seen as a 1-brane. However, for a large number of coinciding
D1 branes, this can be a wide and nearly flat D3 brane. This ``condensation''
of D1 branes to form a D3 brane is the dielectric effect observed in
\cite{Myers9910} (see also \cite{ARS0003}). 

The structure of this work is as follows. In the next section, the
supersymmetry of the system is considered and a differential equation for
the shape of the D3 brane is derived. In section \ref{s-BPS}, the
solutions of this equation are analyzed and in the last section, the
quantization of the parameter $\psi_0$ is discussed.

\newsection{Preserved Supersymmetry}
\secl{s-SUSY}

Type II string theory in flat space-time, has supersymmetry parametrized by
two Weyl-Majorana spinors $\xi_L,\xi_R$ (originating, respectively,
from the left-moving and right-moving sectors).
In the type IIB theory, both spinors have the same chirality and can be
combined to a single complex Weyl spinor
\beql{cxi} \xi=\xi_L+i\xi_R \hs. \eeq
In the conventions used in this work, the chirality condition is
\beql{Chiral} \wbar\Gm\xi=-\xi \hs, \eeq
where
\beql{Gm-def} \{\Gm_A,\Gm_B\}=2\eta_{AB} \hsc \eta=\diag\{-,+,\ldots,+\} \hs,
\eeq 
\[ \Gm_A\dg=\Gm_0\Gm_A\Gm_0 \hs, \]
\beql{BGm-def} \Gm_{A_1\ldots A_r}=\Gm_{[A_1}\ldots\Gm_{A_r]} \hsc
\wbar\Gm=\Gm_{0\ldots9} \hs. \eeq
The Majorana condition is
\beql{Maj-def} (\xi_{L,R})_c=\xi_{L,R} \hs, \eeq
where $\xi\goto\xi_c$ is ``charge-conjugation'':
\beql{conj-def} \xi_c=D\xi^* \hsc D\inv\Gm_A D=-\Gm_A^* \hsc D^*D=1 \hs
\eeq
(so, for $\xi$ in eq. (\ref{cxi}), $\xi_c=\xi_L-i\xi_R$.)

In a background including branes and/or bulk fields, at least some of this
supersymmetry is broken. In this section we identify the supersymmetry that
remains unbroken in the background considered in this work.
First we consider the supersymmetry preserved by the NS5 background
and then, the further influence of D3 branes is analyzed.

\subsection{The NS5 Geometry}

The geometry (metric $ds^2$ in the string frame, dilaton $\Phi$ and
NS 3-form field strength $H$) induced by $k$ coinciding flat NS5 branes is
\cite{CHS9112}
\begin{eqnarray}
\label{NS5ds} ds^2 & = & dx^2+hdy^2 \hs, \\
\label{NS5Phi} e^{2\Phi} & = & g_s^2h\hs \hs, \\
\label{NS5H} H & = & -*_4dh \hspace{1cm}
(\hs H_{klm}=-\ep_{klmn}\pt_nh\hs), 
\end{eqnarray}
where $x^\mu=(x^0,x^1,x^2,x^3,x^4,x^5)$ parametrize the directions parallel to
the NS5 branes, $y^m=(y^6,y^7,y^8,y^9)$ correspond to the transverse
directions and $g_s$ is the string coupling far from the brane.
At this stage we consider coinciding branes, for which $h$ is the following
harmonic function on the transverse space
\beql{h-def} h=1+\frac{kl_s^2}{r^2} \hsc r=|y|=\sqrt{k}l_se^\phi \eeq
(where $l_s$ is the string coupling).
In this case, eqs. (\ref{NS5ds}),(\ref{NS5H}) can be rewritten as
\begin{eqnarray}
\label{NS5ds-sph} ds^2 & = & dx^2+hr^2(d\phi^2+d\Om_3^2) \hs, \\
\label{NS5H-sph} H & = & 2kl_s^2\om_3
\end{eqnarray}
\internote{$\om_3=\frac{1}{r^3}*_4dr=-\frac{1}{2}*_4(\frac{1}{r^2})$}
where $d\Om_3^2$ and $\om_3$ are the metric and volume form on the unit
3-sphere $S^3_{6789}$.
Far from the branes (for $r\gg\sqrt{k}l_s$), $h\approx1$ and the geometry is
trivial: Minkowski space with a constant $\Phi$ and a vanishing $H$.
Close to the brane (for $r\ll\sqrt{k}l_s$), $hr^2\approx kl_s^2$ and the
geometry is that of a throat
\beql{throat} \RR^{5,1} \times \RR_\phi \times S^{3}_{6789} \hs, \eeq
with an $H$ flux through $S^3_{6789}$ and a dilaton $\Phi$ depending
linearly on $\phi$. 
In most of this work, this geometry will be considered semi-classically.
For this to be a justified approximation, one must assume:
\begin{itemize}
\item Small string coupling (loop expansion parameter: $e^\Phi\ll1$:

this is always violated for $r$ close enough to the NS5 branes;
however, for $g_s$ sufficiently small, the strongly-coupled region can be
pushed arbitrarily deep into the throat.
\internote{$e^\Phi=1$ means $r=\sqrt{k}g_sl_s$}
\item Small curvature:

this is always true outside the throat; the throat is also weakly-curved
whenever $k\gg1$.
\end{itemize}

We look for the supersymmetry transformations that preserve this geometry,
\ie, for which the variation of all the fields vanishes. Since all the
fermionic fields vanish, the only non trivial conditions come from the
variation of the fermionic fields - the complex Weyl spinor $\lm$ and the
(complex) gravitino $\psi_M$. The resulting equations are \cite{Schwartz83}:
\internote{Comments:
\nlb In the present conventions: $\wbar\Gm\lm=\lm$, $\wbar\Gm\psi_M=-\psi_M$.
\nlb The following equations are for vanishing RR fields.}
\begin{eqnarray}
\label{SUSY-lm} 0\hs = & \dl\lm &
=\hs i\left(P_M\Gm^M\xi_c-\frac{1}{24}G_{NKL}\Gm^{NKL}\xi\right) \hs, \\
\label{SUSY-psi} 0\hs = & \dl\psi_M & =\hs D_M\xi
+\frac{1}{96}\left(g_{MM'}G_{NKL}\Gm^{M'NKL}-9G_{MKL}\Gm^{KL}\right)\xi_c \hs,
\end{eqnarray}
where 
\beql{PG-def} P=\frac{1}{2}d\Phi \hsc G=\sqrt{g_s}e^{-\Phi/2}H \eeq
\internote{
\nlb More generally: $P=d\rho$, $G=e^{-\rho}H$,
where $\rho=\half\al(\Phi-\Phi_0)$, $\al=\pm1$.
\nlb $G=h^{-\al/4}H$, so $\dl\lm=0$ leads to 
$\xi_c=\al h^{(1-\al)/4}\Gm_{6789}\xi$.
\nlb For $\al=-1$, $\dl\psi_\mu=0$ cannot be satisfied
(the powers of $h$ do not match).}
and $g_{MN}$ is the metric in the {\em Einstein} frame, and is related to the
metric $g^{st}_{MN}$ in eq. (\ref{NS5ds}) by
\[ g_{MN}=\sqrt{g_s}e^{-\Phi/2} g^{st}_{MN} \hs. \]
The covariant derivative of a spinor is
\beql{Der-def} D_M\xi=\left(\pt_M+\frac{1}{4}\Om_{MKL}\Gm^{KL}\right)\xi \hsc
\eeq
\[ \Om_{MKL}=\hat\Om_{KLM}+\hat\Om_{KML}+\hat\Om_{MLK} \hsc
\hat\Om_{MKL}=e^A_L\pt_{[M}e_{K]A} \hs, \]
where $\{e^A=e^A_Mdx^M\}$ is an orthonormal frame
\[ g_{MN}=\eta_{AB}e^A_Me^B_N \]
and the matrices $\Gm^M$ are related to those in eq. (\ref{Gm-def}) by
\[ \Gm_A=\eta_{AB} e^B_M \Gm^M \hs. \]
\internote{Clarification:
\nlb In a general background, a spinor is a section in a vector bundle with
the principle bundle being that of orthonormal frames.
Its transformation is, therefore, correlated with that of the orthonormal
frame ($\in Pin(9,1)$) and it is {\em invariant} under coordinate changes
($\in$Diff).
\nlb Below we use the orthonormal frame $\{h^{-1/8}dx^\al, h^{3/8}dy^a\}$.}
For the geometry (\ref{NS5ds})-(\ref{NS5H}),
eqs. (\ref{SUSY-lm}), (\ref{SUSY-psi}) simplify to
\beql{SUSY-lm-psi} 0=\dl\lm\sim\xi_c+\Gm_{6789}\xi \hsc
0=\dl\psi_\mu=\pt_\mu\xi \hsc 0=\dl\psi_m\sim\pt_m(h^{1/16}\xi) \hs, \eeq
and the general solution is
\beql{SUSY-NS5} \xi=\xi_0h^{\frac{1}{16}} \hsc \Gm_{6789}\xi=-\xi_c \hs, \eeq
where $\xi_0$ is a constant spinor.
\internote{Details: see Appendix \ref{App-SUSY}.}

\subsection{D3 Branes}
\secl{SUSY-D3}

Next we consider D3 branes in the above background. The supersymmetry
preserved by such a brane configuration depends on the charges it carries.
D branes carry RR charges. A D3 brane carries a charge $Q_3$, coupling to the
RR 4-form $C_{[4]}$ and, in some situations (as the present one),
also a charge $Q_1$ (``D1 charge''), coupling to the RR 2-form $C_{[2]}$.
This is expressed by the low-energy effective action of the D3 brane
(see for example \cite{Pol98})
\beql{SD3}
S=g_sT_3\left[-\int d^4\zt e^{-\Phi}\sqrt{-\det(G_{\mu\nu}+\Fc_{\mu\nu})}
+i\int(C_{[4]}+\Fc\wedge C_{[2]})\right] \hs,
\eeq
where
\beql{Tp-def} T_p=\frac{1}{g_sl_s(2\pi l_s)^p} \eeq
is the tension of a D$p$ brane in a trivial background,
$G_{\mu\nu}$ is the bulk metric (pulled-back to the worldvolume of the brane)
and $\Fc$ is a 2-form living on the brane such that $d\Fc=H$ is the bulk NS
3-form (also pulled back to the  worldvolume of the brane).
The 2-form $\Fc$ is conventionally decomposed as
\beql{Fc-dec} \Fc=B+2\pi l_s^2 F \hs, \eeq
where $B$ is the bulk Kalb-Ramond field (with $dB=H$) and $F$ is a $U(1)$
gauge field strength on the brane (with $dF=0$)%
\ftl{B-gauge}{Note that the action (\ref{SD3}) is well defined only when
$H$ is exact in some neighborhood of the D brane.
Note also that the decomposition (\ref{Fc-dec}) is not unique.
The freedom to choose different such decompositions is the gauge freedom
associated with the 2-form potential $B$.}.
\internote{Remarks:
\nlb Validity of the effective action:
\nls neglecting derivatives $\goto$ assuming slow variation of fields
\nls in particular, ``flat'' enough brane.
\nlb The gauge transformation associated with $B$ is
parametrized by a one form $\lm$:
\nl $\dl B = 2\pi l_s^2 d\lm \hsc \dl F = -d\lm$.
\nlb $\Fc=\half\Fc_{\mu\nu}d\zt^\mu\wedge d\zt^\nu$}

To find the preserved supersymmetry in the presence of such charges,
we consider first the simpler situation of D branes in flat space and then
return to the case of interest.

\subsubsection{D Branes in Flat Space}

Consider type IIB string theory compactified on a (rectangular) 3-torus
$T^3_{123}$ with static $N_3$ D3 branes and $N_1$ D1 branes wrapped on the
torus, the D1 branes extended in the $x^1$ direction and delocalized evenly
in the other directions of the torus
(so that translational symmetry is preserved).
The supersymmetry algebra in this situation is (see for example \cite{Pol98}
\beql{SUSY} \{\hat{Q},\hat{Q}\dg\}=-2\left(
\begin{array}{cc} M & (M_3\bt_{123}+M_1\bt_1)\Gm_0 \\
\Gm_0\dg(M_3\bt_{123}+M_1\bt_1)\dg & M \end{array}\right) \hsc
\bt_A=\Gm_A\wbar\Gm \hs,
\eeq
where $\hat{Q}=(Q_L,Q_R)$ are the supercharges,
$M$ is the total mass of the state and
\[ M_p=N_pT_pV_p \]
is the mass of $N_p$ coinciding D$p$ branes of volume $V_p$.
A preserved supersymmetry in a given state corresponds to a spinor
$\hat{\xi}=(\xi_L,\xi_R)$ ($\xi_L,\xi_R$ being two Weyl-Majorana spinors
with $\wbar\Gm=-1$) for which $\hat{Q}\dg\hat\xi$ vanishes in that state
and this implies that $\hat\xi$ is an eigenvector of the matrix on the r.h.s.
of eq. (\ref{SUSY}) with a vanishing eigenvalue.
The resulting ``BPS condition'' is
\beql{BPS-flat} M\xi=i(M_3\Gm_{0123}\xi+M_1\Gm_{01}\xi_c) \hsc
\xi=\xi_L+i\xi_R \hs. \eeq
``Squaring'' eq. (\ref{BPS-flat}), one obtains that the mass of such a BPS
configuration must be
\beql{Mass-flat} M=\sqrt{M_3^2+M_1^2} \hs. \eeq
\internote{Details: (recall that $\wbar\Gm\xi=-\xi$)
\nlb $\Gm_0\dg=-\Gm_0$, $\Gm_I\dg=\Gm_I$ $\Goto$
\nl $[(M_3\bt_{123}+M_1\bt_1)\Gm_0]=\Gm_{01}(-M_3\Gm_{23}+M_1)\wbar\Gm$,
$\hs[\ldots]\dg=\Gm_{01}(M_3\Gm_{23}+M_1)\wbar\Gm$,
\nlb $\Gm_A\dg\Gm_A=1$, $\Gm_{23}\dg=-\Gm_{23}$ $\Goto$
$M^2\xi=[\ldots][\ldots]\dg\xi=(M_3^2+M_1^2)\xi$.}
Note that this is less then the sum of the masses of the constituents, so this
is a real (non-threshold) bound state of the D3 and D1 branes%
\footnote{Performing T-duality in the $x^2$ direction, one obtains two types
of D2 branes, extended in the (13) and (12) directions respectively.
In this situation, the BPS bound state is naturally identified as a
configuration with parallel D2 branes with orientation
$dx^1\wedge(M_3 dx^3+M_1 dx^2)$.}.

The condition (\ref{BPS-flat}) is an integrated version of a corresponding
local condition. Because of the homogeneity of the present configuration, this
local condition can be identified simply by dividing by the volume of the
torus. The result is
\beql{BPS-local} T\xi=iT_3(N_3\Gm_{0123}\xi+n_1\Gm_{01}\xi_c)
\hsc T=T_3\sqrt{N_3^2+n_1^2} \hs, \eeq
where
\beql{den-def} \frac{n_1}{(2\pi l_s)^2}=\frac{N_1}{V_{23}} \eeq
is the ``number density'' of the D1 branes.

Finally, consider a configuration with a single D3 brane, in which the D1
branes are replaced by a 2-form field $\Fc$ on the worldvolume of the D3
brane, with the same charges and preserved symmetries as above.
Translational symmetry implies that $\Fc$ is constant.
To identify its value, we consider the resulting coupling to the
RR 2-form $C_{[2]}$. {}From the D3 brane action (\ref{SD3}) one obtains
\[ ig_sT_3\int_{D3}\Fc\wedge C_{[2]} \]
and this should be equated with the coupling of $N_1$ D1 branes in the $x^1$
direction
\[ iN_1g_sT_1\int_{D1} C_{[2]} \hs. \]
This implies
\beql{Fc-D1} \Fc=n_1\om_{23} \hs, \eeq
where $\om_{23}$ is the volume form on the 2-torus $T_{23}^2$ transverse to
the D1 branes and $n_1$ is related to the density of the D1 branes by
eq. (\ref{den-def}).

We are ready now to return to the D3 branes in the NS5 background

\subsubsection{D3 Branes in the NS5 Brane Geometry}

We consider a configuration that is static, preserves the $SO(3)_{789}$
symmetry and in which the D3 brane is orthogonal to the NS5 branes
(\ie\ its worldvolume is restricted to fixed $x^i$, $i=1,\ldots,5$). 
In this context, it is useful to use also cylindrical coordinates
($z,\rho,\th,\vph$)
\beql{cyl} (y^6,y^7,y^8,y^9)
=(z,\rho(\cos\th,\sin\th(\cos\vph,\sin\vph))) \hs \eeq
($\th\in[0,\pi]$; $\vph\sim\vph+2\pi$) and spherical coordinates,
replacing $(z,\rho)$ by $\phi\in\RR,\psi\in[0,\pi]$:
\beql{sph} (z,\rho)=r(\cos\psi,\sin\psi) \hsc r=\sqrt{k}l_s e^\phi \hs. \eeq
In these coordinates (see eqs. (\ref{NS5ds-sph}),(\ref{NS5H-sph})),
\begin{eqnarray}
\label{ds-sph}
ds^2 & = & dx^2+h(dz^2+d\rho^2+\rho^2 d\Om_2^2)
\hs=\hs dx^2+hr^2(d\phi^2+d\psi^2+\sin^2\psi d\Om_2^2)\hs, \\
\label{H-sph} H & = & 2kl_s^2\frac{\rho^2}{(z^2+\rho^2)^2}(zd\rho-\rho dz)
\wedge\om_2 \hs = \hs 2kl_s^2\sin^2\psi d\psi\wedge\om_2 \hs,
\end{eqnarray}
\internote{$\om_3=\sin^2\psi d\psi\wedge\om_2$, $r^2d\psi=zd\rho-\rho dz$}
where $d\Om_2^2$ and $\om_2$ are the metric and volume form on the unit
2-sphere $S^2_{789}$
\[ d\Om_2^2=d\th^2+\sin^2\th d\vph^2 \hsc \om_2=\sin\th d\th\wedge d\vph \hs.
\]

In the configuration of the above type, the D3 brane is wrapped on
$S^2_{789}$ and is extended in an additional direction
\[ t\propto\sin\chi dz + \cos\chi d\rho \hs, \]
where $\chi$ is independent of the $S^2_{789}$ coordinates.
Because of the rotational symmetry, it is enough to consider $y^8=y^9=0$,
$y^7\ge0$ and there $t\propto\sin\chi dy^6 + \cos\chi dy^7$ ,
so the non-vanishing components of $t$ are
\beql{chi-def} (t_6,t_7)=(\sin\chi,\cos\chi) \hs. \eeq
We assume that the 2-form $\Fc$ on the D brane is proportional to $\om_2$
(and, therefore, $SO(3)_{789}$-symmetric)
\beql{Fc-sph} \Fc=kl_s^2 f(\psi)\om_2 \hs. \eeq
Integrating $d\Fc=H$, one obtains%
\footnote{Identifying $B=[B_0+kl_s^2(\psi-\frac{1}{2}\sin2\psi)]\om_2$
(see eq.(\ref{Fc-dec})), this corresponds to the case of a
purely-magnetic field-strength on the D brane: $F=F_0\om_2$
($B_0+2\pi l_s^2 F_0=-kl_s^2\psi_0$).}
\beql{f-def} f(\psi)=\psi-\psi_0-\frac{1}{2}\sin2\psi \hs. \eeq
\internote{ $f'(\psi)=2\sin^2\psi$ $\Goto$ $d[f\om_2]=2\om_3$.}
The above configuration is very similar to the configuration in flat space
considered before: a D3 brane wrapped on a homogeneous compact 2D surface $S$
(a 2-sphere in the present case), with an $\Fc$ field proportional to the
volume form on $S$.
The BPS condition is, therefore, eq. (\ref{BPS-local})
translated to the present notation:
\beql{BPS-NS}
\sqrt{1+n_1^2}\,\,\xi=i(t_a\Gm_{0a89}\xi+n_1t_a\Gm_{0a}\xi_c) \hs, \eeq
where $t_a$ is defined in eq. (\ref{chi-def}). To identify $n_1$, one notes
that the radius of the 2-sphere, on which the D3 brane is wrapped, is
\beql{R2} R_2=\rho\sqrt{h}=l_s\sqrt{k(1+e^{2\phi})}\sin\psi \hs, \eeq
so the volume form is $R_2^2\om_2$ and, comparing eq. (\ref{Fc-sph}) to
(\ref{Fc-D1}), one obtains
\beql{Nc-NS} n_1=\Fc_{\hat\th\hat\vph}=\frac{f}{g} \hsc
g=\frac{\rho^2h}{kl_s^2}=(1+e^{2\phi})\sin^2\psi \hs.\eeq

We will consider a D3 brane which, far from the NS5 brane, is flat, extended
in the (789) directions and has a vanishing $n_1$. The restriction on
preserved supersymmetry imposed by this part of the brane is
\beql{BPS-D3} \xi=i\Gm_{0789}\xi \hs. \eeq
\internote{Comment:
\nl This equation includes a choice of orientation that
corresponds below to $|\chi|<\pi/2$.}
Combining this with the restriction imposed by the NS5 brane
(eq. (\ref{SUSY-NS5})), one obtains
\beql{BPS-D1} \xi=-i\Gm_{06}\xi_c \eeq
(which is the restriction imposed by a D1 brane extended in $x^6$).
Substituting eqs. (\ref{BPS-D3}),(\ref{BPS-D1}) in the BPS condition
(\ref{BPS-NS}), one obtains
\[ \sqrt{1+n_1^2}\,\,\xi=t_a\Gm_a(\Gm_7-n_1\Gm_6)\xi
=(1+n_1\Gm_{67})e^{\chi\Gm_{67}}\xi \]
(where in the second equality one uses $\Gm_{67}^2=-1$ and the explicit
expression (\ref{chi-def}) for $t_a$). $\Gm_{67}$ anti-commutes with
$\Gm_{0789}$, therefore, this equation can be compatible with eq.
(\ref{BPS-D3}) only if it is an identity, which can be written as
\beql{BPS-id} 1+n_1\Gm_{67}=\sqrt{1+n_1^2}\,\, e^{-\chi\Gm_{67}} \hs. \eeq
This is equivalent to $\arg(1-in_1)=\chi$ and, therefore, also to
\[ \tan\chi=-n_1 \hs. \]
Finally, using the definition (\ref{chi-def}) of $\chi$ and the expression
(\ref{Nc-NS}) for $n_1$, the condition for
preserved supersymmetry translates to the following differential equation
\beql{BPS-cyl} \frac{dz}{d\rho}=-\frac{f}{g} \eeq
and, in spherical coordinates%
\footnote{Using $r^2d\phi=zdz+\rho d\rho$, $r^2d\psi=zd\rho-\rho dz$.},
\beql{BPS-sph} \frac{d\phi}{d\psi}=\frac{\Ac_0}{\Ac_1} \hs, \eeq
\begin{eqnarray*}
\Ac_0 & = & g\sin\psi-f\cos\psi
=\sin\psi-(\psi-\psi_0)\cos\psi+e^{2\phi}\sin^3\psi \hs, \\
\Ac_1 & = & g\cos\psi+f\sin\psi
=(\psi-\psi_0)\sin\psi+e^{2\phi}\sin^2\psi\cos\psi \hs.
\end{eqnarray*}

It is constructive to compare this equation to the corresponding BPS equation
in a trivial background. It can be reproduced from eq. (\ref{BPS-cyl}) by
considering a very large distance $r$ from the NS5 branes. This means that
$h\approx1$ and $f$ (which is a function of $z/\rho$) is approximately
constant, representing a closed $\Fc$: $\Fc=\Fc_0\om_2$. The resulting 
equation is \cite{CM9708}
\[ \frac{dz}{d\rho}=-\frac{\Fc_0}{\rho^2} \hs.\]
Note that, for a D3 brane that extends over regions with different $z/\rho$,
the configuration of the brane may differ significantly from that in a
trivial background, even if the brane is everywhere far from the NS5
branes.

We end this section with a comment on antibranes.
NS5 branes and antibrane differ in the sign
of the 3-form field strength $H$ (eq. (\ref{NS5H})).
D3 branes and antibranes differ in the sign of their
RR charges (the second term in eq. (\ref{SD3})). In both types of branes,
this leads to a sign change in the equation for the preserved supersymmetry
(eqs. (\ref{SUSY-NS5}) and (\ref{BPS-D3}) respectively), but there is
{\em no change} in the differential equation (\ref{BPS-cyl}), so in all cases
the D3 brane has the same shape.

\newsection{BPS D3 Branes}
\secl{s-BPS}

In this section we analyze the solutions of the BPS equation (\ref{BPS-cyl})
\beql{BPS-sc} \frac{dz}{d\rho}=-\frac{kl_s^2f}{\rho^2h}
=-\frac{1}{1+e^{2\phi}}\cdot\frac{f(\psi)}{\sin^2\psi} \hs.\eeq
Such an analysis was performed, for a D5 brane in the geometry induced by D3
branes, in \cite{CGS9810} and the results are qualitatively the same.
This should be expected, since the configurations are related by duality.

\subsection{The Shape of the Brane}
\secl{s-prof}

The solutions of eq. (\ref{BPS-cyl}) have the following properties:
\begin{itemize}
\item
The BPS equation is invariant under $\psi\goto\pi-\psi$ ($z\goto-z$) and
$\psi_0\goto\pi-\psi_0$, so it is enough to consider $\psi_0\ge\frac{\pi}{2}$.
\item
For $\rho\neq0$, $dz/d\rho$ is finite, so the D3 profile is described by
a smooth function $z(\rho)$ in the interval $0<\rho<\infty$.
The sign of its derivative is governed by $f(\psi)$.
\internote{Comments:
\nlb $\rho(z)=0$ is a solution.
\nls It describes a D3 brane wrapped on a vanishing 2-sphere and extended
radially.
\nl Since it has a vanishing tension, this solution is not physical.
\nls For $f_{\rho=0}\neq0$, this is the asymptote for the $\rho\neq0$
solutions.
\nlb For $0\le\psi_0\le\pi$: $\Ac_0>\cos\psi(\tan\psi-\Dl\psi)>0$
\nl $\Goto$ $\psi(r)$ is also a finite smooth
function.
\nls It is increasing iff
$0<\Ac_1=\sin2\psi\cos\psi[e^{2\phi}+\frac{2\Dl\psi}{\sin2\psi}]$.}
\item
For $\rho\goto\infty$: $\frac{dz}{d\rho}\goto0$, so $\psi\goto\frac{\pi}{2}$.
For $\psi_0\neq\frac{\pi}{2}$ this leads to
\beql{z-inf} z\approx z_\infty+\left(\frac{\pi}{2}-\psi_0\right)
\frac{kl_s^2}{\rho}
\eeq
(with $z_\infty$ being an integration constant).
\internote{For $\psi_0=\frac{\pi}{2}$:
$f\approx2(\psi-\frac{\pi}{2})\approx2\frac{z}{\rho}$
and $\sqrt{k}l_s,|z|\ll\rho$
\nl $\Goto$ $z\approx z_\infty e^{kl_s^2/\rho^2}$
$\approx z_\infty(1+\frac{kl_s^2}{\rho^2})$.}
\item
The behavior at $\rho\goto0$ depends on the value of $\psi_0$.
It can be shown that in this limit, $z\goto0$ iff
$0<\psi_0<\pi$.
\internote{Details
\nlb Assuming $z\goto0$ means
$\lim_{\rho\goto0}(\frac{dz}{d\rho}-\frac{f}{\sin^2\psi})=0$
$=\lim_{\rho\goto0}(\frac{dz}{d\rho}-\cot\psi)$
\nl $\Goto 0=\lim_{\rho\goto0}f-\half\sin2\psi=(\psi-\psi_0)$
\nl which is possible only for $0\le\psi_0\le\pi$
\nlb On the other hand, for $0<\psi_0<\pi$, assuming $z\goto z_0>0$,
one obtains finite $h$ and
$\psi\goto0 \Goto f\goto-\psi_0<0 \Goto z\sim-1/\rho$,
which is a contradiction.
\nls An analogous argument holds for $z_0<0$.}
Considering first the case of $z\not\goto0$,
for $\psi_0=\pi$ (implying $f\goto0$) the solution is
\[ z\approx -r_0+\frac{kl_s^2}{3r_0(kl_s^2+r_0^2)}\rho^2 \hs, \]
while for $\psi_0>\pi$, it is
\[ z\approx -\frac{(\psi_0-\pi)kl_s^2}{\rho}+z_0 \]
(with $r_0>0$ and $z_0$ being integration constants).
\internote{Details: 
\nl For $\psi\goto\pi$:
$f=\tilde{\psi}_0-\frac{3}{2}\tilde{\psi}^3+\Oc(\tilde{\psi}^5)$,
$\tilde{\psi}=\pi-\psi\approx-z/\rho$
\nl $\Goto \frac{dz}{d\rho}\approx-\frac{kl_s^2}{kl_s^2+z^2+\rho^2}$%
$\frac{z_2+\rho^2}{\rho^2}(\tilde{\psi}_0+\frac{2}{3}(\rho/z)^3)$.}
\item
When both $z$ and $\rho$ are small (compared to $\sqrt{k}l_s$), the last term
in $\Ac_0,\Ac_1$ (see eq. (\ref{BPS-sph})) is negligible and the BPS
condition becomes
\beql{BPS-throat} \frac{d\phi}{d\psi}=\frac{1}{\psi-\psi_0}-\cot\psi \hs. \eeq
It is solved by
\beql{prof-throat} r=r_0\frac{|\psi-\psi_0|}{\sin\psi} \eeq
(where $r_0$ is an integration constant)
and by $\psi=\psi_0$ (which is the $r_0\goto\infty$ limit of the solution
(\ref{prof-throat})).
\internote{Comments:
\nlb The meaning of the parameters:
$\psi\stackrel{r\goto0}{\goto}\psi_0$,
$\rho\stackrel{\psi\goto0}{\goto}\rho_0|\psi_0|$,
$\rho\stackrel{\psi\goto\pi}{\goto}\rho_0|\pi-\psi_0|$.
\nlb For $0\le\psi_0\le\pi$: $\psi(r)$ is either constant or monotonic.
\nlb For $\psi_0=\frac{\pi}{2}$: $\psi(r)=\frac{\pi}{2}$ ($z=0$)
is an exact solution also outside the throat.}

\end{itemize}
The resulting configurations are illustrated in figures
\ref{f-pe}, \ref{f-pg} and \ref{f-pl}.
They are parametrized by $\psi_0$, which represents the value of $\Fc$ and,
for each such value, there is an additional continuous parameter that can be
identified with $z_\infty$ in eq. (\ref{z-inf}).
Far from the NS5 branes, the configuration is the trivial one%
\footnote{Recall that this was {\em a choice} (see below eq. (\ref{Nc-NS}))
that was used to derive the BPS equation (\ref{BPS-cyl}).}:
a flat D3 brane (at $z=z_\infty$) with a vanishing $\Fc$.
The shape of the brane depends continuously on $z_\infty$
(\ie, the topology does not change), while in the dependence
on $\psi_0$, there is a qualitative change at $\psi_0=\pi$
(and similarly at $\psi_0=0$):
\begin{itemize}
\item $\psi_0=\pi$ (figure \ref{f-pe}):
\begin{figure}
\pct{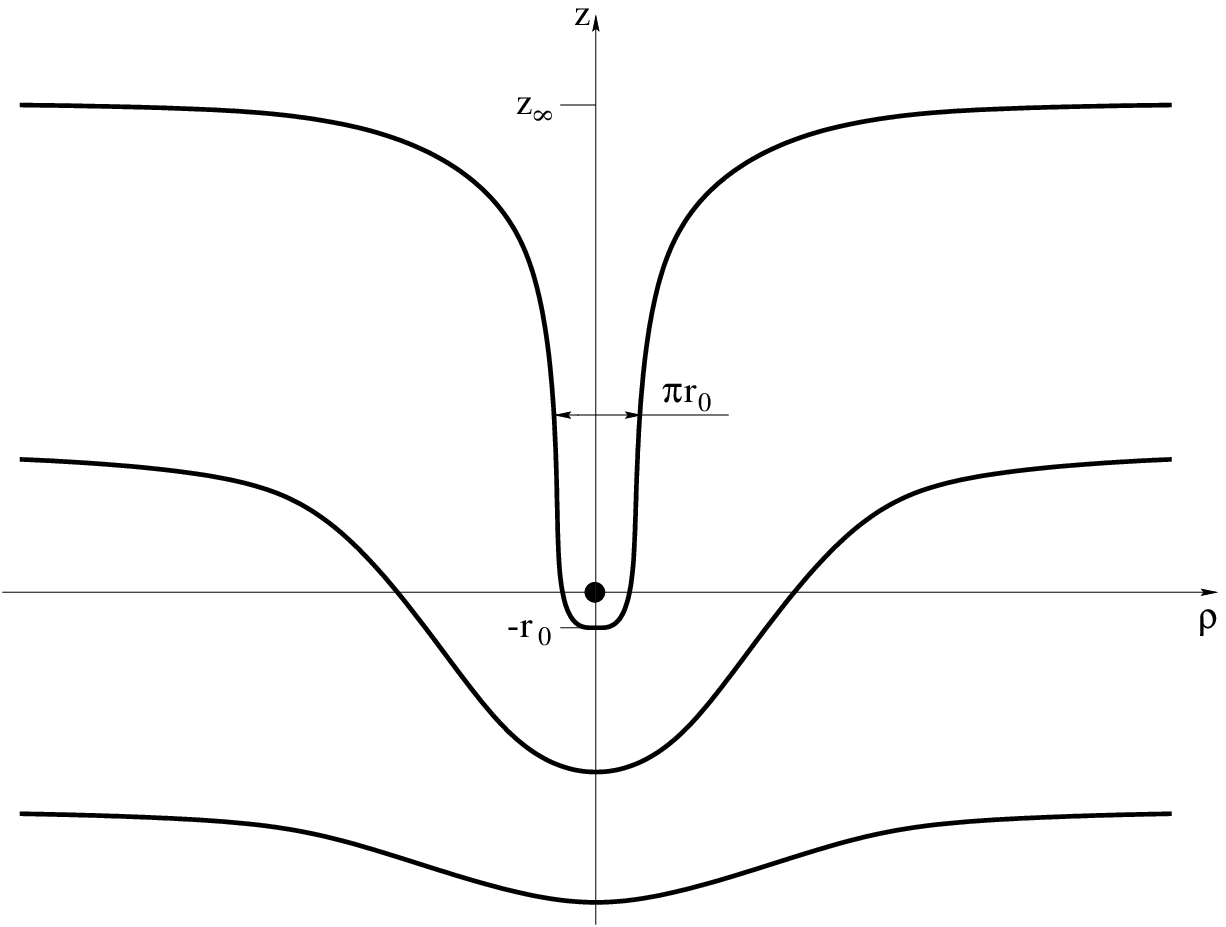}
\capl{f-pe}{D3 brane profiles with $\psi_0=\pi$}
\end{figure}

For $z_\infty$ negative and large enough, the whole brane is nearly flat;
as $z_\infty$ is increased, the NS5 branes repel the D3 brane and, as a
result, a (finite) tube is formed.
\item $\psi_0>\pi$ (figure \ref{f-pg}):
\begin{figure}
\pct{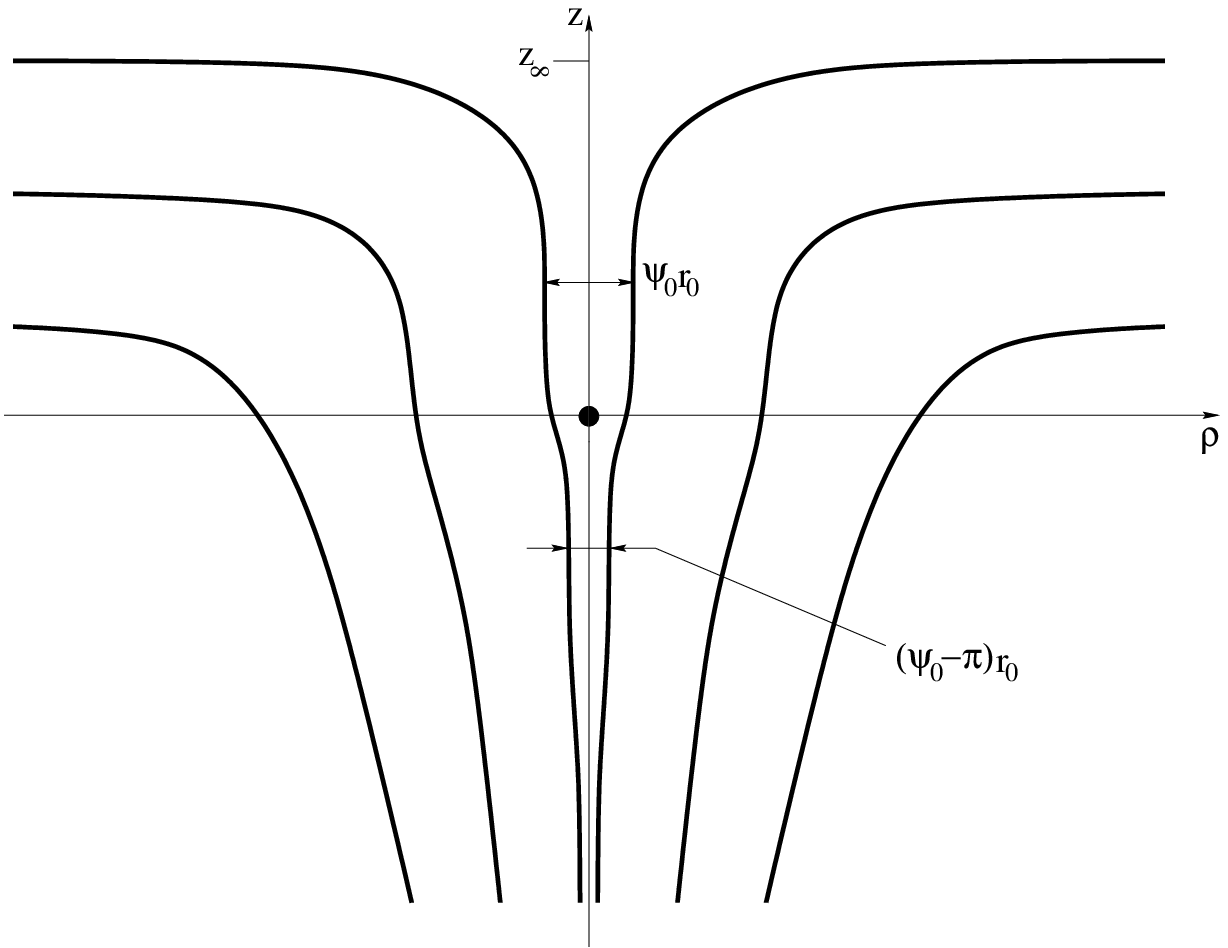}
\capl{f-pg}{D3 brane profiles with $\psi_0>\pi$}
\end{figure}

For any $z_\infty$, the D3 brane avoids the $\rho=0$ line%
\ftl{f-sing}{ In fact, the D3 brane avoids the $\rho=0$ for any
$\psi_0\neq0,\pi$. This means that the brane avoids the singularities of
$\Fc$. These singularities can be deduced from the components of $\Fc$ in
an orthonormal frame. The non-vanishing component
$\Fc_{\hat\th\hat\vph}$ is given in eq. (\ref{Nc-NS}),
so $\Fc$ is singular at $\psi=0$ (for $\psi_0\neq0$) and at $\psi=\pi$
(for $\psi_0\neq\pi$).}
\internote{Comment:
\nlb This is a result of the SUSY requirement and {\em does not} follow from
energy considerations.
\nlb For example, far from the NS5 branes (and with $kl_s^2=1$):
\nls $g=\rho^2$, $f=-\psi_0$ and extremization of energy
(eq. (\ref{E-ex-cyl})) gives $\frac{z'\Ac}{\sqrt{{z'}^2+1}}=c=$const.,
which means $z'=\frac{c}{\sqrt{\Ac^2-c^2}}$.
\nls The BPS equation $z'=\frac{\psi_0}{\rho^2}$ chooses $c=\psi_0$
and this gives a divergent $z$ at $\rho=0$.
\nls However, for $c^2\neq\psi_0^2$, $z$ is finite at $\rho=0$!}%
and an infinite tube is formed, extending to $z\goto-\infty$.
\item $0<\psi_0<\pi$ (figure \ref{f-pl}):
\begin{figure}
\pct{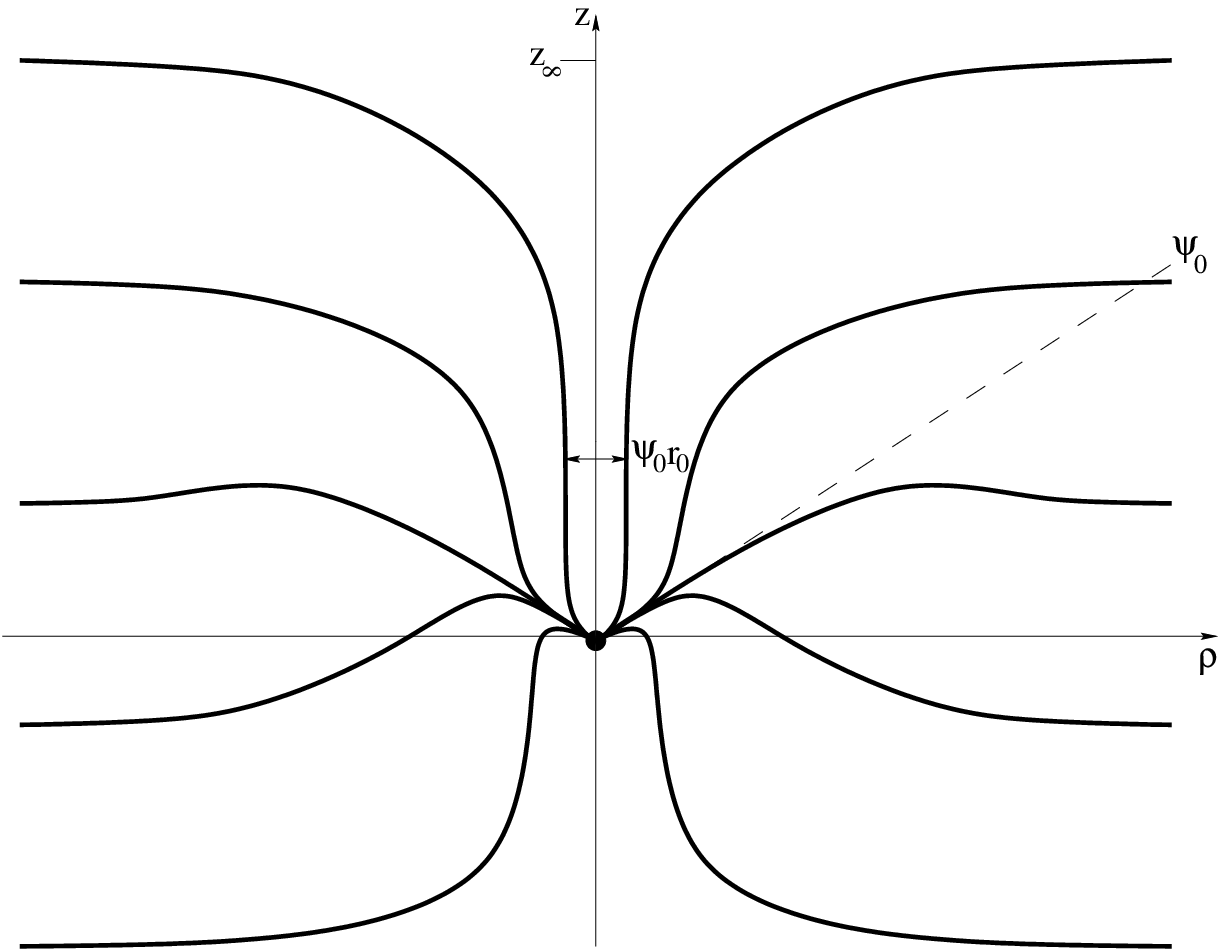}
\capl{f-pl}{D3 brane profiles with $0<\psi_0<\frac{\pi}{2}$}
\end{figure}

These configurations also include an infinite tube, but this time it 
extends to $\phi\goto-\infty$ in the throat, reaching the NS5 branes.
In this case, $\psi_0$ has a clear geometrical meaning: deep in the
throat, the D3 brane approaches a cylinder with $\psi=\psi_0$
(independently of $z_\infty$),
so this is the direction from which the D3 brane approaches the NS5 branes.
\end{itemize}
For any $\psi_0$, by changing $z_\infty$ in the appropriate direction,
the tube can be made arbitrarily narrow.
When the radius $R_2$ of the tube (eq. (\ref{R2})) becomes
comparable to $l_s$ (or smaller),
the semi-classical description used here breaks down and
the tube is more appropriately interpreted as a 1-brane.
The resulting brane configuration, as seen at large scales, is composed of
a flat D3 brane and a 1-brane ending on the D3 brane, both branes being
orthogonal to each-other (and to the NS5 branes).
In the configuration in figure \ref{f-pg}, the 1-brane is semi-infinite,
while in the other two cases, the 1-brane has a finite extension,
its lower end being at the NS5 branes.
The nature of this 1-brane will be discussed in the next section.

\subsection{The Energy of the Brane}
\secl{s-en}

The low-energy effective action for the D3 brane is given in eq. (\ref{SD3}).
For a static brane, it takes the form $S=-E\Dl x^0$, where
$E$ is the energy of the brane (measured w.r.t. the time coordinate $x^0$).
In the present configuration, the energy is a functional of $z(\rho)$
\beql{ED3-cyl}
E=\frac{k}{\pi}T_1\int d\rho\sqrt{{z'}^2+1}\,\,\Ac \hsc
\Ac^2=g^2+f^2 \eeq
(where $g,f$ are given in eqs. (\ref{Nc-NS}),(\ref{f-def}) respectively
and the $z'=dz/d\rho$)
\internote{Details:
\nl $ds^2=-(dx^0)^2+h[({z'}^2+{\rho'}^2)d\zt^2+\rho^2 d\Om_2^2]$
\nl $\Goto$ $E=\int d\zt\cdot 4\pi\cdot g_s T_3\cdot(g_s^2 h)^{-1/2}$
$\cdot\{h({z'}^2+{\rho'}^2)[(h\rho^2)^2+(kl_s^2f)^2]\}^{1/2}$
\nl where the prime denotes differentiation w.r.t. the coordinate $\zt$,
parametrizing the location of the D3 brane in the $(z,\rho)$ plane.}
and the equation obtained from extremizing it is
\beql{E-ex-cyl}
\frac{d}{d\rho}\left[\frac{z'\Ac}{\sqrt{{z'}^2+1}}\right]
=\sqrt{{z'}^2+1}\frac{\pt\Ac(z,\rho)}{\pt z} \hs.
\eeq
One can show that any solution of the BPS condition (\ref{BPS-cyl})
also exrtemizes the energy (\ie\ solves eq. (\ref{E-ex-cyl}))%
\footnote{One way to do this is as follows: using
\[ \Ac^2=g^2+f^2 \hsc z'=-\frac{f}{g} \hs, \]
eq. (\ref{E-ex-cyl}) simplifies to
\[ \frac{\pt f}{\pt\rho}=-\frac{\pt g}{\pt z} \]
(where $f,g$ are considered as function of $z,\rho$) and one can verify that
$f,g$ given in eqs. (\ref{Fc-sph}),(\ref{f-def}) indeed satisfy this last
equation.}.
\internote{Details: $\frac{\Ac}{\sqrt{{z'}^2+1}}=g$, $\hs z'g=-f$}
\internote{Stabilization of a wrapping on a non-minimal surface:
\nl (for a D2 brane in the throat [BDS0003])
\nls The tension wants to decrease $\sin\psi$
\nls When $\Fc$ depends on $\psi$ (this is possible only when $H\neq0$)
it can have a contra effect.
\nls Indeed, in the present case, $\Ac$ has a unique minimum at $\psi=\psi_0$
\nl (for $0<\psi_0<\pi$, the points $\psi=0,\pi$ are extremal points but not
minima or maxima).}

\newsection{Quantization}
\secl{s-quant}

So far, the discussion was purely classical. The quantum theory induces
quantization restrictions on the parameters appearing in the classical
description. These can be derived in various ways and lead to identical
results.

\subsection{Consistency of the Fundamental String Interactions}

The quantization restrictions can be derived from the coupling of the
fields $H$ and $\Fc$ to the fundamental string
(this was analyzed in \cite{KS9609} and considered further in
\cite{AS9812}\cite{Gawedzki9904}\cite{Stanciu9909}\cite{Pawel0003}%
\cite{EGKRS0005}\cite{KKZ0005}\cite{Stanciu0006}). 
Consider such a string, propagating in a spacetime $\Mc$. Its worldsheet
$\Sg\subset\Mc$ may have a boundary $\pt\Sg$ which, however, is necessarily
confined to the worldvolume $D\subset\Mc$ of a D brane.
Assuming that there exists a three-dimensional
manifold $\hat\Sg\in\Mc$ such that%
\footnote{The existence of such $\hat\Sg$ for any $\Sg$ is expressed
mathematically as the triviality of the relative homology group
$H_2(\Mc/D,\ZZ)$.
In the present case, $H_2(\Mc,\ZZ)=H_1(D,\ZZ)=0$, which indeed implies
$H_2(\Mc/D,\ZZ)=0$.
The more general case $H_1(D,\ZZ)\neq0$ was considered in \cite{KS9609}.}
$\pt\hat\Sg=\Sg+\Sg_D$, $\Sg_D\subset D$,
\internote{Comments:
\nlb $\pt\Sg$ may have different parts (corresponding to different D branes),
with {\em different} $\Fc$ (at the same place in $\Mc$).
\nlb An exact sequence (of {\em integral} homology groups):
\nl $\ldots H_3(\Mc/D)\goto H_2(D)\goto H_2(\Mc)\goto H_2(\Mc/D)$
$\goto H_1(D)\ldots$
\nl $C_3\goto C_2\goto C_2$, $\Sg+\Sg_D\goto\Sg+\Sg_D$, $\Sg\goto\pt\Sg$}
the coupling of the string to the fields $H$ and $\Fc$ is
\beql{Sint} S_{\rm Int}=\frac{1}{2\pi l_s^2}\left(\int_{\hat\Sg}H
-\int_{\Sg_D}\Fc\right) \hs. \eeq
Note that in topologicly-trivial situations, when
\beql{BA} H=dB \hsc F\equiv\frac{\Fc-B}{2\pi l_s^2}=dA \hs, \eeq
eq. (\ref{Sint}) reduces to the following familiar form:
\[ S_{\rm Int}=\frac{1}{2\pi l_s^2}\int_\Sg B+\int_{\pt\Sg}A \hs. \]

The phase $e^{iS_{\rm Int}}$ (appearing in the path integral) must be
independent of the choice of $\hat\Sg$ and this translates to
\beql{quant} \frac{1}{2\pi l_s^2}\left(\int_{C_3}H-\int_{C_2}\Fc\right)
\in 2\pi\ZZ \hs, \eeq
for any $C_3\subset\Mc$ with $\pt C_3=C_2\subset D$.
In particular, for a closed $C_3$ (which always the case for a closed string)
the condition is
\beql{H-int} \frac{1}{(2\pi l_s)^2}\int_{C_3}H \in \ZZ \hs, \eeq
which is an integrality condition on the cohomology class of $H$.
\internote{This is a non-trivial condition only when $H_3(\Mc,\ZZ)$ is
non-trivial.}
For a given $H$ satisfying the integrality constraint (\ref{H-int}),
the further condition (\ref{quant}) with an open $C_3$ is a condition on
the worldvolume $D$ of the D brane and the field $\Fc$ on it.
Note that $C_3$ and $C'_3$ with the same boundary $C_2$ lead to the same
condition (because of the integrality condition (\ref{H-int})) so,
for a given $C_2\subset D$, one can choose a convenient $C_3$.
When the Kalb-Ramond field $B$ is sufficiently regular, so that $C_3$ can be
chosen to avoid the singular locus of $B$ (which means that $H$ is exact
in $C_3$: $H=dB$), condition (\ref{quant}) simplifies
(using the decomposition (\ref{Fc-dec})) to
\beql{F-int} \frac{1}{2\pi}\int_{C_2}F\in\ZZ \hs. \eeq
This form for the integrality condition was proposed recently in
\cite{BDS0003} (see also \cite{Pawel0003}\cite{KKZ0005}\cite{Stanciu0006}),
in the specific case of $\Mc=S^3$, and here this proposal is confirmed.
As pointed out in \cite{BDS0003}, one should question the meaning of an
integrality condition for a quantity like $F$, that is not gauge invariant
(see footnote \ref{B-gauge}). This issue is clarified in the present
derivation: to obtain the above condition on $F$, a (regular) gauge choice of
$B$ is necessary.
\internote{
\nlb The naive guess for an integrality condition would be a condition for
$\int_{S^2}\Fc=4\pi kl_s^2(\tilde{f}-\psi_0)$.
\nlb Mathematically, if $F$ is interpreted as a curvature of a $U(1)$
connection, then $\frac{1}{2\pi}\int F$ is the first Chern class of the
corresponding $U(1)$ bundle, which is an integral class
$\Goto$ the integrality condition (\ref{F-int}) is implicit in any
theory in which $F$ is a $U(1)$ gauge field strength.
\nlb Physically, for a spacelike $C_2$, the integer in eq. (\ref{F-int})
is the magnetic monopole number.}

We now apply the above considerations to the NS5 background. The topology is
$\RR^7\times S^3$. The first factor is trivial homologicly and $H,\Fc$
(in eqs. (\ref{NS5H-sph}),(\ref{Fc-sph})), do not depend on it, so one can
replace $C_3$ by its projection on the second factor, which is a subset of
$S^3$. The only closed $C_3$ is $S^3_{6789}$ itself, so the integrality
constraint (\ref{H-int}) is
\beql{H-int-S3} \frac{1}{(2\pi l_s)^2}\int_{S^3}H \in \ZZ \hs. \eeq
Using eq. (\ref{NS5H-sph}) and $\int\om_3=2\pi^2$, this translates to
$k\in\ZZ$.
As to an open $C_3$, we choose for the bulk Kalb-Ramond field
\beql{B-sph} B=kl_s^2(\psi-\frac{1}{2}\sin2\psi)\om_2 \hs, \eeq
which is singular only at $\psi=\pi$ (see footnote \ref{f-sing}).
For $C_2$ avoiding%
\footnote{$C_2$ passing through $\psi=\pi$ can be deformed away from the
singular point and from the results below one can see that this case
does not lead to additional restrictions.}
$\psi=\pi$, one can choose $C_3$ that also avoids the singular point and use
the integrality condition (\ref{F-int}). 
Using the explicit expressions for $B$ and $\Fc$
(eqs. (\ref{B-sph}),(\ref{Fc-sph}),(\ref{f-def})), one obtains
\beql{F-sph} F=-kl_s^2\psi_0\om_2 \hs, \eeq
which is independent of $\psi$, so one can project $C_2$ on $S^2_{789}$,
obtaining an integral multiple of $S^2_{789}$.
The integrality condition (\ref{F-int}) is, therefore
\[ \frac{1}{2\pi}\int_{S^2}F\in\ZZ \hs, \]
which leads to
\beql{psi-int} \psi_0=N\frac{\pi}{k} \hsc N\in\ZZ \hs. \eeq
Note that, although the derivation used a choice of gauge, the result
(\ref{psi-int}) has a gauge invariant meaning, since $\psi_0$ determines
the field $\Fc$ (see eq. (\ref{f-def})).
Note also that, while the condition (\ref{F-int}) is a condition on $F$, the
dynamics  (\ie, the effective action (\ref{SD3}) turned it to a restriction
on the possible configurations of the D3-brane (as described in the
previous section). In particular, for the configuration $\psi=\psi_0$ in the
throat region ($r\ll\sqrt{k}l_s$; see the comment below eq.
(\ref{prof-throat})), there are only $k-1$ possible 2-spheres
on which the D3 brane can be wrapped, corresponding to
$N=1,\ldots,k-1$ (this was first derived in \cite{AS9812}).
\internote{Comment [AS9812]: the above quantization condition
\nl $\frac{k}{\pi}\left(\int_{C_3}\om_3-\int_{C_2}\om_2\right)\in 2\pi\ZZ$
(with $\om_3=d\om_2$) is an extension of
the standard Bohr-Sommerfeld quantization condition
$\frac{k_c}{\pi}\int\om_2\in 2\pi\ZZ$ (?).}

\subsection{The Integrality of the Number of Branes}
\secl{s-int-brane}
Although the integrality of the parameter $k$ was obtained in the previous
subsection using a consistency argument, this integrality is obviously also
a consequence of the identification of $k$ as the number of NS5 branes.
Likewise, as will be shown in this subsection, the integrality condition
(\ref{psi-int}) on $\psi_0$ may be derived from the integrality of the
number of D1 branes that can end on the D3 brane.

As described in the previous section, for $\psi_0>0$ and $z_\infty$ large
enough, the configuration includes a long and thin tube.
When the radius $R_2$ (eq. (\ref{R2})) of this tube is small relative to
$l_s$, it should be understood semi-classically as a 1-brane.
To identify this brane, it is instructive to rewrite eq.  (\ref{ED3-cyl})
for the energy in the following form%
\footnote{These expressions are derived using the BPS equation $z'=-f/g$,
and the relation $rdr=(zz'+\rho)d\rho$.}
\beql{ED3BPS} E=\frac{k}{\pi}T_1\int d\rho\frac{\Ac^2}{g}=T_1\int N'_Tdr
\hsc N'_T=\frac{k}{\pi}\frac{\Ac^2}{|\Ac_0|} \hs.  \eeq
Comparing this to the energy of a (static) D1 brane,
extended in the radial direction (at a fixed point in $S^3_{6789}$)%
\footnote{This expression is derived using the analog of eq. (\ref{SD3}) for
a D1 brane. Note that the factors of the harmonic function $h$ cancel.}
\beql{ED1} E=T_1\int dr \hs, \eeq
one sees that the above 1-brane can be identified as a collection of
$N'_T$ D1 branes. Such an identification requires that $N'_T$ in eq.
(\ref{ED3BPS}) should be an integer. To find the resulting condition
on $\psi_0$, we consider $\psi_0>0$ and $z_\infty\goto\infty$.
This corresponds to $r_0\goto0$ in eq. (\ref{prof-throat}), so for a fixed
$r$ in the throat (satisfying $0<r\ll\sqrt{k}l_s$), $\psi\goto0$,
which implies
\[\Ac\approx\Ac_0\approx-f\approx\psi_0 \hs. \]
This gives
\beql{N1-id} N'_T=\frac{k\psi_0}{\pi} \hs, \eeq
leading to the integrality condition (\ref{psi-int}) with $N=N'_T$.
The identification of the tube as a D1 brane was obtained here considering
the tension of the tube.
Alternatively, one can consider the D1 charge of the tube and compare it
with that of a D1 brane. This was done in subsection \ref{SUSY-D3}.
Eqs. (\ref{den-def}),(\ref{Fc-D1}),(\ref{Fc-sph}) give
\[ N'_Q=\frac{1}{(2\pi l_s)^2}\int_{S^2} \Fc =\frac{k}{\pi}f(\psi) \]
which, for $\psi\goto0$ coincides with eq. (\ref{N1-id})%
\footnote{More precisely, one obtains $N'_Q=-\frac{k\psi_0}{\pi}$. The minus
sign means that the D1 branes are oriented in the {\em negative} direction of
$z=x^6$.}.
\internote{The D1 orientation:
\nlb (\ref{Nc-NS}): in the $f\hat{r}$ direction
\nlb (\ref{BPS-cyl}): $f\hat{r}$ means $-\hat{z}$.}

Therefore, a distant observer sees $N'_T$ coinciding D1 branes (extended
in the $z=x^6$ direction) ending on a flat D3 brane (extended in the
$(x^7,x^8,x^9)$ directions). For $\psi_0\le\pi$, the number of branes is
at most $k$ and all of them have finite extent, the other end being on the
NS5 branes. For $\psi_0>\pi$, the number $N'_T$ of D1 branes above the
NS5 branes $(z>0)$ is larger then $k$, but not all of them end on the NS5
branes, since the D3 tubes extends also to $z<0$.
Comparing the charge and tension of the tube above and below the NS5 branes
(note the difference in the width of the tubes in figure \ref{f-pg}),
one finds exactly $k$ D1 branes ending on the NS5 branes, while the other
$N'_T-k$ are semi-infinite, extending from the D3 brane to $z\goto-\infty$.
As will be seen in the next subsection, such a configuration can be obtained
from a configuration with $N'_T=k$ by moving some of the NS5 branes down to
$z\goto-\infty$.

As the width of the tube is increased,
both $N'_T$ and $N'_Q$ change continuously and are no longer integers.
This is particularly bothering for $N'_Q$, which represents a RR charge.
This apparent puzzle, observed in \cite{BDS0003}, was resolved in
\cite{Taylor0004} (see also \cite{AMM0005}\cite{Maldorf0006}).
Using the decomposition (\ref{Fc-dec}) with the choice (\ref{B-sph}) of $B$,
one obtains
\[ N'_Q=N_Q+\dl N_Q \hsc N_Q=\frac{1}{2\pi}\int_{S^2}F \hsc
\dl N_Q=\frac{1}{(2\pi l_s)^2}\int_{S^2}B=\frac{1}{(2\pi l_s)^2}\int_{C_3}H
\hs, \]
where $C_3$ is the disc $[0,\psi]$ in $S^3_{6789}$ (in which $B$ is regular!).
It was shown in \cite{Taylor0004} that, as the width of the tube is increased,
there is a bulk contribution to the RR charge which exactly cancels $\dl N_Q$,
so the total RR charge remains
\beql{N1-int}  N_Q=\frac{1}{2\pi}\int_{S^2}F=-\frac{k\psi_0}{\pi} \hs, \eeq
which is integral and equal to the number of D1 branes ending on the D3
brane.
\internote{Comments:
\nls For $\psi_0=\pi$ and narrow tube, the brane configuration can be viewed
as a D3 wrapping the $S^3$ with $N_Q=k$ D1's ending on it.
\nls In this situation, The D1 charge of $k$ branes is needed to cancel the
D1 charge induced on the $S^3$ part by the $H$ flux of the NS5's
(the CS term in the D3 brane effective action) [BDS0003].
\nls When the tube has a finite width, part of the $H$ flux escapes,
so the amount of {\em needed} D1 charge carried by the tube changes,
as does the {\em actual} charge.}

\subsection{A Dynamical Explanation}

Both the above derivations of the restriction (\ref{psi-int}) on the possible
D3 brane configurations use indirect arguments and do not provide a
real explanation for the restriction. Such an explanation will be given
in this subsection. To obtain it, one separates the NS5 branes along the $z$
axis and studies the resulting D3 brane configuration near $\rho=0$.
Considering all classical solutions with $0<\psi_0<\pi$
(\ie, ignoring the quantization restriction (\ref{psi-int})),
one finds two types of configurations:
\begin{enumerate}
\item the D3 brane passes between two NS5 branes;
\item the D3 brane intersects one of the NS5 branes.
\end{enumerate}
Then, considering the integrality condition (\ref{psi-int}),
one observes that it precisely distinguishes between these two types,
allowing only the first one%
\footnote{This relation between the quantized $\psi_0$ and the possible
positions of the D brane among the NS5 branes was suggested independently
in \cite{EGKRS0005}.}.
This suggests the following dynamical origin for the quantization
restriction:
\begin{quote}\em
The D3 and NS5 branes (in the relative orientation considered here)
repel each-other and avoid intersecting one-another.
\end{quote}
Recall that this repulsion was explicitly observed in the solution with
$\psi=\pi$ (figure \ref{f-pe}).
The only situation in which the two branes touch one-another is when NS5
branes from both sides of the D3 brane are brought to coincidence, trapping
the D3 brane between them.

We now give some details of the analysis.
The geometry induced by the above distribution of NS5 branes is still given
by eqs. (\ref{NS5ds})-(\ref{NS5H}),
but this time with the following harmonic function
\[ h=1+\sum_i\frac{kl_s^2}{|\vec{y}-\vec{y}_i|^2} \hs, \]
where $\{\vec{y}_i=(z_i,0,0,0)\}$ are the locations of the branes.
The supersymmetry preserved by the geometry is still given by
(\ref{SUSY-NS5}), since in its derivation, the explicit form
(\ref{h-def}) of $h$ was not used.
Likewise, in the derivation of the BPS equation (\ref{BPS-cyl}),
only the $SO(3)_{789}$ symmetry of $h$ (and $f$) was used (as manifested by
the use of cylindrical coordinates), so it remains valid also
in the present case. 
\internote{Details:
\nlb The $SO(4)_{6789}$-symmetric $h$ (\ref{h-def}) of coinciding NS5
branes was used only in:
\nls The expression for $H$ (\ref{H-sph})
\nls The expression for $\Fc$ (and $f$) (\ref{Fc-sph})
\nls The l.h.s. of eqs. (\ref{ds-sph}) and $\Ac_0,Ac_1$ in (\ref{BPS-sph})}
The function $f$ appearing in this equation is not that given in eq.
(\ref{f-def}), but a more general function of $z,\rho$.
To obtain it, one recalls eqs. (\ref{Fc-sph}),(\ref{NS5H}):
\[ kl_s^2df\wedge\om_2=d\Fc=H=-*_4dh \hs, \]
which imply a linear relation between $df$ and $dh$ and, since $dh$ is a
superposition of the contributions from each NS5 brane, so is $df$.
Therefore, $kf=\sum_i f_i$, where $f_i$ is obtained from the function $f$
in eq. (\ref{f-def}) by shifting it to the location of the $i$'th NS5 brane.
In particular, $f$ is locally constant along the $z$ axis, and decreases
discontinuously by $\frac{\pi}{k}$ at each of the locations of the NS5 branes.

To study the solutions of the BPS equation,
we start with coinciding NS5 branes and a D3 brane configuration
corresponding to $0<\psi_0<\pi$ and a large positive $z_\infty$. We recall
from the previous subsection that this configuration includes a thin 1-brane
with tension
\beql{Ten} T=N'_T T_1 \hsc N'_T=\frac{k\psi_0}{\pi} \hs. \eeq
Now we separate the NS5 branes along the $z$ axis
until there is a large distance between them, so that 
in any given region, there is at most one NS5 brane that has a significant
influence on the geometry. Thus, one can use
the results of subsection \ref{s-prof} to study the corresponding D3 brane
configuration.
Changing the NS5 brane distribution means, technically, changing the functions
$h$ and $f$. 
Recall that $f$ is defined up to an integration constant. We fix this
ambiguity by requiring that the part of the D3 brane far above the NS5 branes
remains unchanged. This is achieved by holding fixed the value $-\psi_0$ of
$f$ on the $z$ axis, above all the NS5 branes and, consequently, the tension
of the 1-brane above the NS5 branes is still given by (\ref{Ten}).
Now one follows this solution down, to the region of the NS5 branes.
Near the upper NS5 brane there are the following possibilities:
\begin{itemize}
\item $(0<)N'_T<1$:
the D3 brane intersects the NS5 brane, as in figure \ref{f-pl}; the angle of
approach is $N'_T\pi$.
\item $N'_T=1$:
the tube closes just below this NS5 brane, as in figure \ref{f-pe}, but
above the other NS5 branes;
\item $N'_T=1$:
the tube continues down below the NS5 brane, with a reduced tension,
corresponding to $N'_T$ replaced by $N'_T-1$ in eq. (\ref{Ten}).
Note that this means that exactly one D1 brane ends on the NS5 brane.
\internote{This is a reproduction of the ``s rule''}
\end{itemize}
In the last case, one can continue to follow the tube to the next NS5 brane
and so on. Denoting
\[ N'_T=\hat{N}-1+\frac{\hat{\psi}_0}{\pi}
\hspace{1cm} (\hat{N}=1,\ldots,k,\hs 0<\psi\le\pi) \hs, \]
one obtains that when $N'_T$ is integer ($N'_T=\hat{N}$, $\hat{\psi}_0=\pi$),
the D3 brane passes between the $\hat{N}$th and $(\hat{N}+1)$th NS5 brane,
while otherwise, the D3 brane intersects the $\hat{N}$th NS5 brane,
with $\hat{\psi}_0$ being the approach angle.
These are the results stated in the beginning of the subsection.

Strictly speaking, one should bear in mind that the throat of a
single NS5 brane is highly curved, so to use semi-classical consideration,
one should avoid these regions. This can be done by considering tubes that
are wide enough. For such tubes, $N'_T$ is not simply related to $\psi_0$,
but one can consider, instead, $N_Q$ (defined in eq. (\ref{N1-int})), which
is not modified by changing the width.
Alternatively, one can restrict attention to NS5 distributions where there
are several concentrations of branes, each with many branes, leading to a
geometry which is weakly curved everywhere. Considering only such situations,
one can show that configurations in which the D3 brane passes between NS5
branes always correspond to $\psi_0$ which satisfies the integrality
condition.

\subsection{Boundary Conditions in the Worldsheet CFT} 

So far, the analysis was semi-classical, the D3 brane being represented by
a classical manifold in the NS5 geometry.
In the throat region of this geometry ($r\ll\sqrt{k}l_s$), one can do better.
The corresponding (supersymmetric) 2D non-linear $\sg$ model is an exactly
solvable Conformal Field Theory (CFT) \cite{CHS9112},
so one can consider the full quantum dynamics of the fundamental string.
This results with the same quantization conditions on $k$ and $\psi_0$.
We now review this approach and its relation to the previous ones.

In the worldsheet formulation of the string dynamics,
a D brane corresponds to a boundary of the worldsheet
and different branes are distinguished by different boundary conditions
imposed at the corresponding boundaries
(see \cite{BPPZ9908} and references therein).
The boundary conditions studied so far for the throat background
(\ref{throat}) have a factorized form
(\ie\ do not mix the various factors in (\ref{throat})).
Geometrically, this means that the intersection of the D brane with the
3-sphere is independent of the other factors. This is the case only in a
subset of the D3 brane configurations that were identified semiclassically
in the previous section: $\psi_0$ is restricted to the range $(0,\pi)$
and for each such $\psi_0$, $z_\infty$ has a specific value, corresponding to
$r_0\goto\infty$ in eq. (\ref{prof-throat}), so that in the throat the brane
is described by $\psi=\psi_0$. Therefore, in the CFT approach one considers
only this subset of D3 branes.

The $S^3_{6789}$ factor in (\ref{throat})) can be identified
as the $SU(2)$ group manifold, and the corresponding CFT is
(the supersymmetric extension of) $SU(2)$ WZW model.
The $SO(4)_{6789}$ symmetry of the geometry is manifested by the existence,
in the CFT, of an $\what{su}(2)$ affine algebra generated, in the left/right
sectors, by the ``currents'' $J,\tilde{J}$ respectively%
\footnote{The level of this algebra is $k$ and its representation theory
implies that $k$ is a (positive) integer, in agreement with the previous
considerations.}.
The preservation of the $SO(3)_{789}$ symmetry by the D brane is achieved by
imposing a ``gluing condition'' $J=\tilde{J}$, as part of the boundary
condition.
\internote{Details:
\nlb HW irreps have finite dimensional irreps of $SU(2)$
``pseudo-spin'', implying that $k_B$ is an integer [Polchinski,II67].
\nlb The gluing condition $J=\tilde{J}$ means that the boundary condition
preserves the diagonal affine $\what{su}(2)$.
\nlb The choice of the point $\psi=0$ as the identity element of the group
identifies the diagonal $SU(2)$ with $SO(3)_{789}$.}
This gluing condition leads  \cite{AS9812}%
\footnote{The geometrical meaning  of gluing conditions as above was also
discussed in \cite{KO9612}\cite{ST9805}\cite{Stanciu9901}\cite{FFFS9909}%
\cite{Stanciu9909}\cite{FS9909}\cite{FS0001}\cite{EGKRS0005}\cite{KKZ0005}%
\cite{Stanciu0006}.}
(as required by the symmetry) to a worldvolume that intersects the
3-sphere at a 2-sphere with a fixed $\psi=\psi_0$.
Imposing also the preservation of supersymmetry and following the
approach in \cite{Cardy89}, one can show that the possible boundary conditions
are in one-to-one correspondence with the primary states of the bosonic
affine algebra%
\footnote{The affine algebra acts both on the bosons
and fermions in the CFT, with levels $k-2$ and $2$ respectively.}
and there are $k-1$ such states, labeled by their $SU(2)$ spin $j$:
$2j=0,\ldots,k-2$.
\internote{Details: see Appendix \ref{App-Group}.}
This agrees with the number of allowed 2-spheres
implied by the quantization condition (\ref{psi-int}) (as observed in
\cite{AS9812}), suggesting the identification
\beql{psi-j} \psi_0=N\frac{\pi}{k} \hsc N=2j+1=1,\ldots,k-1 \hs. \eeq
\internote{Comments:
\nlb [AS9812] derived the quantization condition on $\psi_0$, using the
approach of KS9607.
\nls Unlike here, they considered the bosonic WZW and identified $k=k_B$,
therefore, two of the boundary states were identified with the
{\em degenerate} 2-spheres $\psi_0=0,\pi$.
\nlb $N$ is the dimension of the spin $j$ representation.
Is this related to the identification of this brane as a condensation
of $N$ D1 branes? (are the two $SU(2)$ groups related?)}
Evidence for this identification was found recently in \cite{FFFS9909},
where the expectation values of bulk fields were calculated and their
source (identified as the D-brane) was found to be concentrated around a
2-sphere with $\psi=\psi_0$ given by eq. (\ref{psi-j}). 
It is worth emphasizing that the analysis in \cite{FFFS9909} is performed in
the framework of the {\em exact} CFT and is, therefore, valid also for $k$
which is not large (although the D brane is found to be delocalized
in a band of width $\Dl\psi\sim2\frac{\pi}{k}$, which is significant for
small $k$).
\internote{Details
\nls The vev of the metric is a class function (depends only on $\psi$).
\nls The peaks are at $\psi=\psi_0$ given by eq. (\ref{psi-j}).
\nls The width (between zeros) $\Dl\psi=2\frac{\pi}{k}$.}

Note that the values $\psi=0,\pi$ ($N=0,k$) are absent from the list
(\ref{psi-j}),
although they are allowed by the quantization condition (\ref{psi-int}).
These values correspond to a D1 brane ending on the NS5 branes and,
therefore, the absence of a corresponding boundary condition in the
worldsheet description may seem disturbing.
This puzzle is, however, resolved by the discussion in subsection
\ref{s-int-brane}. There it was demonstrated that a D3 brane with
$\psi_0=\frac{\pi}{k}$ (corresponding to $j=0$ in eq. (\ref{psi-j})),
wrapped on a 2-sphere with a sufficiently
small radius should be identified as a D1 brane. In the present case
(\ie, in the throat; with $\psi=\psi_0$), the radius of this cylinder is
\[ R_2=l_s\sqrt{k}\sin\frac{\pi}{k}\siml\frac{\pi}{\sqrt{k}}l_s\siml l_s \]
(see eq. (\ref{R2})), so it is indeed small enough.
\internote{Comments:
\nlb In this sense, D1 branes are seen by NS5 branes as cylindrical D3 branes.
\nlb A configuration corresponding to a semi-infinite D1 brane ending on the
NS5 branes corresponds, in the throat to a {\em non-factorized} boundary
condition ($\psi$ depends on $r$).
\nls The identification of this boundary condition is an
interesting problem, left for future work.}

\vspace{1cm}
\noindent{\bf Acknowledgment:}
\nl I wish to thank D. Kutasov for helpful discussions.
This work is supported in part by DOE grant \#DE-FG02-90ER40560.


\appendix
\renewcommand{\newsection}[1]{
 \vspace{10mm} \pagebreak[3]
 \refstepcounter{section}
 \setcounter{equation}{0}
 \message{(Appendix \thesection. #1)}
 \addcontentsline{toc}{section}{
  App. \protect\numberline{\Alph{section}}{\hs\hs\boldmath #1}}
 \begin{flushleft}
  {\large\bf\boldmath Appendix \thesection. \hspace{5mm} #1}
 \end{flushleft}
 \nopagebreak}


\ifinter\beginsup
\input dbisup
\endsup\fi


\end{document}